\documentclass[aps,pre,superscriptaddress,amsmath,amssymb,twocolumn]{revtex4}
\usepackage{graphicx}
\usepackage{amsmath}
\usepackage{hyperref}
\usepackage{color}
\usepackage{mathrsfs}
\usepackage{calrsfs}
\definecolor{darkblue}{RGB}{8,81,156}
\hypersetup{colorlinks,breaklinks,
            linkcolor=darkblue,urlcolor=darkblue,
           anchorcolor=darkblue,citecolor=darkblue}

\definecolor{dark-red}{RGB}{215,48,39}

\date{\today}

\allowdisplaybreaks

\begin{document}

\title{Thermodynamic and Kinetic Anisotropies in Octane Thin Films}

\author{Amir Haji-Akbari}
%\email{hajakbar@princeton.edu}
\affiliation{Department of Chemical and Biological Engineering, Princeton University, Princeton NJ 08544}

\author{Pablo G. Debenedetti}
\email{pdebene@exchange.princeton.edu}
\affiliation{Department of Chemical and Biological Engineering, Princeton University, Princeton NJ 08544}

\date{\today}

\begin{abstract}
Confinement breaks the translational symmetry of materials, making all thermodynamic and kinetic quantities functions of position. Such symmetry breaking can be used to obtain configurations that are not otherwise accessible in the bulk. Here, we {use computer simulations to} explore the effect of substrate-liquid interactions on thermodynamic and kinetic anisotropies induced by a solid substrate. We consider $n$-octane nano films that are in contact with substrates with varying degrees of attraction, parameterized by an interaction parameter $\epsilon_S$. Complete freezing of octane nano films is observed at low temperatures, irrespective of $\epsilon_S$, while at intermediate temperatures, a frozen monolayer emerges at solid-liquid and vapor-liquid interfaces. By carefully inspecting the profiles of translational and orientational relaxation times, we confirm that the translational and orientational degrees of freedom are decoupled at these frozen monolayers.  At sufficiently high temperatures, however,  free interfaces and solid-liquid interfaces close to loose (low-$\epsilon_S$) substrates undergo 'pre-freezing`, characterized by mild peaks in several thermodynamic quantities. Two distinct dynamic regimes are observed at solid-liquid interfaces.  The dynamics is accelerated in the vicinity of loose substrates, while sticky (high-$\epsilon_S$) substrates decelerate dynamics, sometimes by as much as  two orders of magnitude. These two distinct dynamical regimes have been previously {reported} by us [\href{http://dx.doi.org/10.1063/1.4885365}{J. Chem. Phys.~\textbf{141}: 024506 (2014)}] for a model atomic glass-forming liquid. We also confirm the existence of two correlations-- proposed in the above-mentioned work-- in solid-liquid subsurface regions of octane thin films, i.e.,~a correlation between atomic density and normal stress, and between atomic translational relaxation time and lateral stress. Finally, we inspect the ability of different regions of an octane film to explore the potential energy landscape by performing inherent structure calculations, and observe no noticeable difference between the free surface and the bulk in efficiently exploring the potential energy landscape. This is unlike the films of model atomic glass formers that tend to sample their respective landscape more efficiently at free surfaces. We discuss the implications of this finding to the ability of octane-- and other $n$-alkanes-- to form ultrastable glasses.
\end{abstract}

\maketitle

\section{Introduction\label{section:intro}}

Confinement alters the thermodynamic and kinetic properties of matter. Such changes can partly arise from  quantum effects at the nanoscale ~\cite{MoiseeviNature1996, DanielChemRev2004, MarinicaNanoLetters2012, HalpernNanoLett2015}. But they can also emerge in purely classical systems, simply due to the presence of confinement-induced interfacial regions. On a  fundamental level, confinement breaks the translational isotropy of a bulk material, making all its physical properties functions of position~\cite{HajiAkbariJCP2014}. The extent of such  anisotropy is, however, variable, and can depend on both the thermodynamic conditions (temperature, density, composition)  as well as the nature of the interface(s), such as boundary conditions (free interface vs.~substrate), interactions (attractive, repulsive, inert) and dimensionality (pores, channels, slabs). Regardless, the behavior of  confined matter can deviate significantly from the bulk in many different ways. Examples include the emergence of new phases~\cite{JakobssonNanoLett2003, MolineroJCP2010, GaoJPhysChemB2015}, changes in thermodynamic properties such as  melting and glass transition temperatures~\cite{ChristensonJPCM2001, TaschinEPL2010, RichertARPC2010, PriestleyMacromolecules2011, OguniJPCB2011}, and in kinetic properties such as  nucleation rates~\cite{SumitPNAS2012, HajiAkbariFilmMolinero2014, CoxJCP2015}, viscosities~\cite{BrinkeLangmuir1995, StaffordSoftMatter2009, RafiqJPCB2010, DaleySoftMatter2012}, diffusivities~\cite{BarratEPL1995, deBeerEPL2012,  KrishnamoortiACSNano2013} and elastic constants~\cite{HirvonenJAppPhys1997, StaffordSoftMatter2009, HoffmanPRL2010}.

The translational anisotropy that is induced as a result of confinement can be utilized for modulating the structural and functional properties of materials. A widely known example is  heterogeneous nucleation~\cite{TurnballJCP1970} in which a substrate enhances a particular first-order phase transition in its vicinity by decreasing the associated free energy barriers. On a microscopic level, this  is mediated by the formation of a subsurface region that is more conducive to nucleation than the homogeneous bulk material. The nucleating potency of a substrate is thus related to the microstructure of the material that is in its proximity, i.e.,~the anisotropy that it induces therein~\cite{MurrayChemSocRev2012}. The relationship between the anisotropy of the subsurface region and the facilitated nucleation can be rigorously explained for phase transitions such as hydrophobic evaporation~\cite{SumitPNAS2012, AltabetJCP2014}, but is far more difficult to discern for more complex phase transitions, such as crystallization~\cite{CoxJCP2015}.

A more recent example of the interesting behavior arising in anisotropic systems is provided by  ultrastable glasses obtained by depositing the vapor of a glass-forming liquid onto a cold substrate. By tuning the substrate temperature, it is possible to form glasses that are far more stable than the ordinary glasses obtained by rapidly quenching the liquid~\cite{EdigerScience2007}.  In addition to their superior thermal and mechanical stability, such ultrastable glasses can be structurally anisotropic as evident from refractive index measurements and X-ray scattering~\cite{DalalJPCB2013, ShakeelPNAS2015, AnkitChemMater2015}. Possible structural differences between ordinary and ultrastable glasses are manifest in the different scaling of their heat capacities in the limit of $T\rightarrow0$~K~\cite{RamosPNAS2014}. It has even been suggested that a first-order liquid-liquid transition might exist between the  ordinary and ultrastable amorphous states~\cite{DawsonPNAS2009}. Vapor deposition has been successfully used for making organic~\cite{EdigerScience2007, IshiiCPL2008, GutierrezTA2009, GutierrezPCCP2010, GutierrezJPCL2010, SoudaJPCB2010, SepulvedaPRL2011, IshiiJPCB2012}, polymeric~\cite{PriestleyNatMat2012} and metallic~\cite{SamwerAdvMat2013} ultrastable glasses. There is, however, a significant gap in understanding  why these vapor-deposited glasses are ultrastable. It has been argued that the enhanced mobility of molecules at the vapor-liquid interface enables them to explore the potential energy landscape more efficiently, thereby giving rise to structures that reside far deeper in the potential energy landscape~\cite{EdigerMacromolecules2014}. Such enhanced mobility has been observed in experimental~\cite{BellJACS2003, ZhuEdigerPRL2011, DaleySoftMatter2012, ForestScience2014} and computational~\cite{ShiJCP2011,dePabloJCP2014, HajiAkbariJCP2014, LyubimovJCP2015} studies of thin films. However, no causal relationship has  been unambiguously established between the existence of such accelerated regions and the formation of ultrastable glasses

A systematic way of identifying a possible link between enhanced surface mobility and increased thermodynamic and kinetic stability is to probe thermodynamic and kinetic anisotropies in films of different compositions, and to assess their sensitivity to changes in thermodynamic conditions. It is of particular interest to study the role of substrates in inducing such anisotropies, and in modifying the properties of the resulting glasses. This is a question that has not been extensively studied, especially in regard to the vapor-deposited ultrastable glasses. Two particular parameters that are relevant in this quest are the temperature and the interaction parameter, with the latter quantifying the strength of (attractive) interactions between the substrate and the liquid.  In our earlier publication~\cite{HajiAkbariJCP2014}, we performed a detailed analysis of such anisotropies in  thin films of a model atomic glass-forming liquid~\cite{KobAndersenPRE1995}, and we discovered two distinct dynamical regimes: accelerated dynamics near  loosely attractive substrates, and decelerated dynamics near sticky substrates. We also established correlations between oscillations in density and in normal stress, and between  oscillations in relaxation time and in lateral stress.

In this work, we revisit the main findings of Ref.~\cite{HajiAkbariJCP2014}, but now in the context of molecular thin films of $n$-octane. We choose $n$-octane as a prototypical chain molecule. Alkane thin films are particularly interesting as they can undergo a process known as \emph{surface freezing}~\cite{WuScience1993, OckoPRE1997} in which a frozen layer emerges at the vapor-liquid interface at temperatures exceeding the equilibrium melting temperature. Also, alkanes constitute one of the most important components of  crude oil, and their presence can result in interesting phase separation phenomena that are typically governed by confinement~\cite{HoepfnerLangmuir2013, NasimEF2013, NasimEF2014, NasimEF2015}. It is therefore worthwhile to inspect the microstructure of interfacial regions of alkane films with an eye towards understanding the mechanism(s) and predicting the kinetics of such phase transition processes.

This paper is organized as follows. Section~\ref{section:system} provides details on the computational setup as well as the force-fields employed. Technical details of the simulations performed are given in Section~\ref{section:simulation}. The procedures utilized for computing spatial profiles of thermodynamic and kinetic properties are given in Section~\ref{section:methods:spatial}. Section~\ref{section:results:qualitative} discusses the qualitative behavior of the octane films, particularly, in regard to surface freezing. Anisotropies in thermodynamic and kinetic properties are discussed in Sections~\ref{section:thermo} and~\ref{section:kinetics}, respectively. Finally, Section~\ref{section:conclusions} is reserved for concluding remarks.

\section{Methods}

\begin{figure}
	\centering
	\includegraphics[width=0.2492\textwidth]{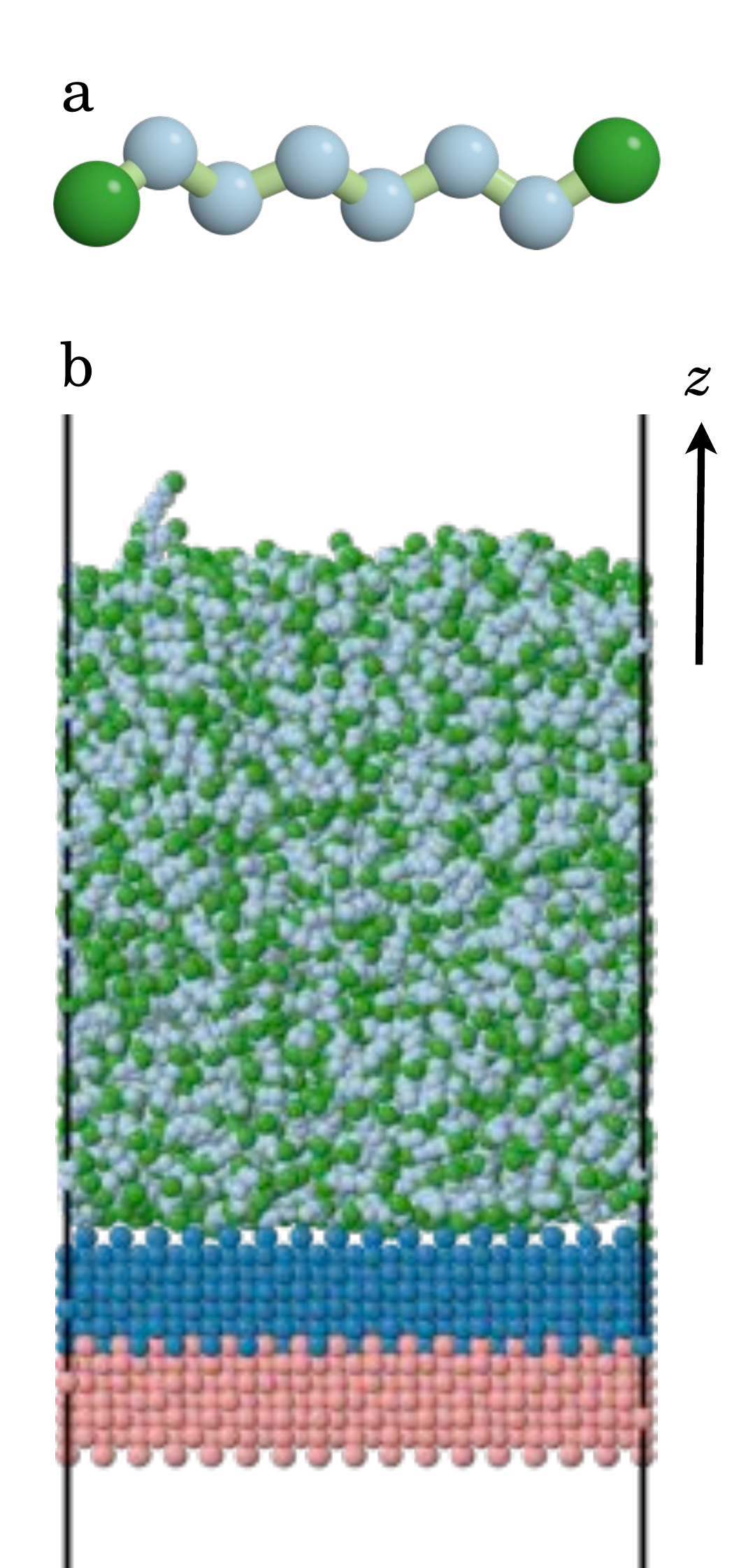}
	\caption{\label{fig:schematic}(a) Representation of an $n$-octane molecule with the NERD force field. Green and light boue sites correspond to the united-atom CH$_3$ and CH$_2$ interaction sites respectively. (b) A liquid film of $n$-octane molecules in the vicinity of an attractive (steel blue) substrate. The light pink substrate is repulsive.}
\end{figure}

\subsection{System Description\label{section:system}}
A schematic representation of the $n$-octane films considered in this work is depicted in Fig.~\ref{fig:schematic}b. Octane molecules are represented with the NERD force-field~\cite{NathDePabloJCP1998} in which each octane molecule is comprised of eight united-atom interaction sites: two for CH$_3$ (green in Fig.~\ref{fig:schematic}a) and six for CH$_2$ (light blue in Fig.~\ref{fig:schematic}a). {Our choice of a united-atom potential such as NERD is due to its simplicity, as we only consider $n$-alkanes as prototypical chain molecules. Indeed, the increased quantitative accuracy obtained from using more realistic, but computationally costly, all-atom force-fields is unlikely to impact the main findings of this work, and can only shift the $n$ values or temperatures for which the reported phenomena are observed.
In the NERD potential}, the non-bonded interactions between these sites are modeled through the Lennard-Jones potential~\cite{LJProcRSoc1924}:
\begin{eqnarray}
V_{LJ}(r) &=& 4\epsilon_{ij}\left[\left(\frac{\sigma_{ij}}{r}\right)^{12}-\left(\frac{\sigma_{ij}}{r}\right)^{6}\right]
\end{eqnarray}
The bonded interactions include  bond-stretching, bond-bending and torsional terms:
\begin{subequations}
\begin{eqnarray}
V_{\text{stretching}}(r) &=& k_s(r-r_0)^2\\
V_{\text{bending}}(\theta) &=& k_b(\theta-\theta_0)^2\\
V_{\text{torsional}}(\phi) &=& \frac12K_1(1+\cos\phi)+\frac12K_2(1+\cos2\phi)\notag\\
&& +\frac12K_3(1+\cos3\phi)
\end{eqnarray}
\end{subequations}
All model parameters are given in Table.~\ref{table:NERD}. Lennard-Jones interactions are shifted and truncated at $r_c=1.38~$nm. 

In addition to the $n$-octane film, there are two substrates in the system: the attractive substrate ($C$ atoms, steel blue  in Fig.~\ref{fig:schematic}b) and the repulsive substrate ($D$ atoms, light pink  in Fig.~\ref{fig:schematic}b). The constituent atoms of both  substrates are arranged into a face-centered cubic (fcc) lattice. They interact with themselves and  the united-atom LJ sites via the Lennard-Jones potential with $\epsilon_{AC}=\epsilon_{AD}=\epsilon_S\epsilon_{AA}$, $\epsilon_{BC}=\epsilon_{BD}=\epsilon_S\epsilon_{BB}$, and $\sigma_{AC}=\sigma_{BC}=\sigma_{AD}=\sigma_{BD}=0.4$~nm. Here, the interaction parameter, $\epsilon_S$, is  used for tuning the strength of attractive (repulsive) interactions between the substrates and  octane molecules. All substrate-octane interactions are truncated and shifted at $r_c=1.2$~nm and $r_c=$0.44~nm for the attractive and repulsive substrates, respectively. The inclusion of a second repulsive substrate is to assure that the octane molecules evaporating from the film would never redeposit onto the opposite side of the substrate as a result of periodic boundary conditions. Simulations are carried out for three distinct values of $\epsilon_S=0.5, 1.0$ and 3.0, with the temperature range $200~\text{K}\le T\le 290$~K considered at each $\epsilon_S$ value. {As will be shown in Section~\ref{section:results}, this temperature range covers the states that are both above and below freezing.}

\begin{table}
	\begin{center}
		\caption{\label{table:NERD}Parameters of the NERD force-field for $n$-octane molecules. A and B correspond to the united-atom CH$_3$ and CH$_2$ sites, respectively.}
		\begin{tabular}{ll}
		\hline\hline
		Parameter~~~~~~~~~~~~~~~~~~~~~~~~~~~& Value~~~~~~~~~~~~~~~~~~~~~~~~~~ \\
		\hline
		$\epsilon_{AA}$ & 0.2066~kcal/mol\\
		$\sigma_{AA}$ & 0.391~nm\\
		$\epsilon_{BB}$ & 0.09106~kcal/mol \\
		$\sigma_{BB}$ & 0.393~nm\\
		$\epsilon_{AB}=\sqrt{\epsilon_{AA}\epsilon_{BB}}$ & 0.1372~kcal/mol \\
		$\sigma_{AB}=(\sigma_{AA}+\sigma_{BB})/2$ & 0.392~nm\\
		$k_s$ & 191.85~kcal/mol$\cdot$\AA$^2$\\
		$r_0$ & 0.154~nm\\
		$k_b$ & 124.26~kcal/mol$\cdot$radian$^2$\\
		$\theta_0$ & 114$^\circ$\\
		$K_1$ & $1.4117$~kcal/mol\\
		$K_2$ & $-0.2711$~kcal/mol\\
		$K_3$ & $3.1465$~kcal/mol\\
		\hline
		\end{tabular}
	\end{center}
\end{table}

\subsection{Simulation Details\label{section:simulation}}
All Molecular Dynamics (MD) simulations are performed using LAMMPS~\cite{PimptonLAMMPS1995} in the isochoric (NVT) ensemble. We integrate Newton's equations of motion using the velocity Verlet algorithm~\cite{SwopeJCP1982} with a time step of 2~fs, and we control temperature using the Nos\'{e}-Hoover thermostat~\cite{NoseMolPhys1984, HooverPhysRevA1985} with a time constant of $\tau=0.2$~ps. All simulations are carried out in cuboidal boxes that are periodic in all dimensions. The simulation box is always longer along the $z$ direction in order to assure the lack of correlation between the liquid film and its periodic images.

In all simulations, the starting configuration is comprised of an $n$-octane film in which all molecules are arranged into a simple cubic lattice. This configuration is initially heated at $T=300$~K for 200~ps in order to melt the crystal. The melted film is then gradually quenched to $T_f$ at a cooling rate of $2.5\times10^{12}$~K/s. The arising configuration is equilibrated at $T_f$ for 4~ns. This equilibration time is far larger than the structural relaxation time in the bulk for the temperatures considered in this work. Throughout the entire process, a separate thermostat is applied to the atoms in the substrate, always maintained at a temperature $T_f$. After this initial equilibration stage, the production runs are carried out for 80~ns.

\subsection{Spatial Profiles\label{section:methods:spatial}}

\subsubsection{Thermodynamic Properties}
The bulk of the methodology that is used for computing the spatial profiles of thermodynamic and kinetic properties is discussed in our earlier publication~\cite{HajiAkbariJCP2014}. For thermodynamic quantities that are time-invariant, spatial profiles can be rigorously determined via simple time averaging of the corresponding property in thin cuboidal slices of the simulation box. Here, profiles of potential energy, atomic and molecular density, lateral and normal stress and lateral radial distribution function are computed in slabs that are $0.025$~nm thick, and the binning of all molecular properties, including molecular density, and radial distribution function, is performed based on the centers of mass of the molecules. We also compute inherent structures using the FIRE algorithm~\cite{BitzekPRL2006}, and the contribution to the average inherent structure potential energy of a slice is from those atoms that were originally located in that slice prior to  energy minimization. We also compute the orientational distribution function (ODF), $f(\theta,z)$, of  $n$-octane molecules defined as follows. First, $\mathcal{G}_i$ the gyration tensor of each $n$-octane molecule is computed from:
\begin{eqnarray}
\mathcal{G}_i &=& \frac{\sum_{j=1}^8m_{i,j}(\textbf{r}_{i,j}-\textbf{r}_{i,CM})(\textbf{r}_{i,j}-\textbf{r}_{i,CM})^T}{\sum_{j=1}^8m_{i,j}}
\end{eqnarray}
Here $m_{i,j}$ and $\textbf{r}_{i,j}$ are the mass and the position of the $j$th united atom  site of the $i$th molecule and $\textbf{r}_{i,CM}$ is the center of mass of the $i$th molecule. Subsequently, $\textbf{v}_i$, the longest principal axis of the $i$th molecule is determined, which is the unit eigenvector corresponding to the largest eigenvalue of $\mathcal{G}_i$. Due to the inversion symmetry of $\mathcal{G}_i$, both $\pm\textbf{v}_i$ will be equally valid choices. In order to avoid any ambiguity, we choose the $\textbf{v}_i$ for which $\textbf{v}_i\cdot\textbf{n}_S\ge0$. Here, $\textbf{n}_S$ is a unit vector perpendicular to the substrate, and pointing towards the liquid.
 The orientational distribution function is then defined as:
\begin{eqnarray}
f(\theta,z) &=& \frac{\left\langle\sum_{i=1}^N\delta\left(|\textbf{v}_i\cdot\textbf{n}_S|-\cos\theta\right)\delta(z_{i,CM}-z) \right\rangle}{\left\langle\sum_{i=1}^N\delta(z_{i,CM}-z) \right\rangle}\notag\\&&\label{eq:ODF}
\end{eqnarray} 
Note that $\int_{0}^{\pi/2}f(\theta,z)\sin\theta d\theta=1$. $f(\theta,z)$ is  utilized for computing further orientational order parameters that are introduced and discussed in Section~\ref{section:results:qualitative}.

\subsubsection{Kinetic Properties}
The time averaging procedure that is used for computing profiles of thermodynamic properties cannot be utilized for kinetic properties such as relaxation times as the latter are obtained from ensemble averages of autocorrelation functions. This difficulty stems from the ambiguity of defining autocorrelation functions in open systems. In our earlier publication~\cite{HajiAkbariJCP2014}, we present a thorough discussion of different heuristics that can be used for defining autocorrelation functions in confined systems.  Here, we take the same convention as used in that work, and define atomic and molecular translational and rotational relaxation times as follows. For translation relaxation, we compute the $z$-dependent self intermediate scattering function:
\begin{eqnarray}
F_{S,X}(q,z,t) &=& \frac{1}{\mathscr{N}_X(z,t)}\left\langle \sum_{i=1}^{N_X}e^{iq||\Delta\textbf{r}_i^{||}(t)||}\mathfrak{D}(z(t),z(0),z) \right\rangle\notag\\
&&\label{eq:sisf}
\end{eqnarray}
Here $X=\text{a, m}$, with a and m corresponding to atomic and molecular self-intermediate scattering functions, respectively. $\Delta\textbf{r}_i^{||}(t)$ is the lateral displacement of entity $i$ over time $t$. $\mathfrak{D}(z_1,z_2,z)=\delta(z_1-z)\delta(z_2-z)$ is an indicator that assures that entity $i$ is present at $z$ both in the beginning and at the end of the time window, and $\mathscr{N}_X(z,t)=\langle \sum_{i=1}^{N_X}\mathfrak{D}(z_i(t),z_i(0),z)\rangle$ is the average number of entities that contribute to the sum for a particular slice in Eq.~(\ref{eq:sisf}). Like thermodynamic properties, contributions to molecular autocorrelation functions are based on the centers of mass of $n$-octane molecules.  For octane films, we use a $q=16~\text{nm}^{-1}$, which corresponds to the first peak of $S(q)$, the structure factor, in the bulk $n$-octane liquid computed in an $NpT$ simulation at $T=300$~K and $p=0$~bar. We observe no significant change in the locus of the maximum of $S(q)$ in bulk octane with temperature. Relaxation times are determined from $F_{S,X}(q,z,t=\tau)=0.2$. The atomic and molecular translational relaxation times are denoted by $\tau_{\text{tr,a}}$ and $\tau_{\text{tr,m}}$, respectively.

In order to quantify molecular rotational relaxation time profiles, we compute the following $z$-dependent orientational auto-correlation function:
\begin{eqnarray}
h_m(z,t) &=& \frac{1}{\mathscr{N}_m(z,t)}\left\langle\sum_{i=1}^{N_m}P_2[\textbf{v}_i(t)\cdot\textbf{v}_i(0)]\mathfrak{D}(z_i(t),z_i(0),z)\right\rangle\notag\\
&&\label{eq:hm}
\end{eqnarray}
with $P_2(\cos\theta)=(3/2)\cos^2\theta-(1/2)$  the second Legendre polynomial. Note that $P_2(\cdot)$ is the natural choice considering the inversion symmetry of the gyration tensor used for determining $\textbf{v}_i$.  Analogously, the relaxation of  orientations of individual bonds is characterized using $h_b(z,t)$, the bond autocorrelation function, defined as follows:
\begin{eqnarray}
h_b(z,t) &=& \frac{1}{\mathscr{N}_b(z,t)}\left\langle\sum_{i=1}^{N_b}P_1[\textbf{b}_i(t)\cdot\textbf{b}_i(0)]\mathfrak{D}(z_i(t),z_i(0),z)\right\rangle\notag\\
&&\label{eq:hb}
\end{eqnarray}
Here $\textbf{b}_i$ is the unit vector in the direction of bond $i$ that connects two united-atom sites on an $n$-octane molecule, and $z_i$ is the $z$ coordinate of the center of the bond. Since a bond is essentially a directed entity with no inversion symmetry, we use $P_1(\cos\theta)=\cos\theta$, the first Legendre polynomial, to quantify its rotational relaxation, as using $P_2(\cdot)$ will lead to loss of relevant orientational information. Similar to translation relaxation times, molecular and bond rotational relaxation times are determined from $h_m(z,t=\tau_{\text{rot,m}})=0.2$ and $h_b(z,t=\tau_{\text{rot,b}})=0.2$, respectively.

Since it is more difficult to obtain suitable statistics for the $z$-dependent autocorrelation functions introduced above, we use slices that are 0.1~nm thick, four times thicker than the slices used for computing $z$-dependent time averages of thermodynamic properties

\section{Results and Discussion\label{section:results}}

\begin{figure*}
	\begin{center}
		\includegraphics[width=0.8250\textwidth]{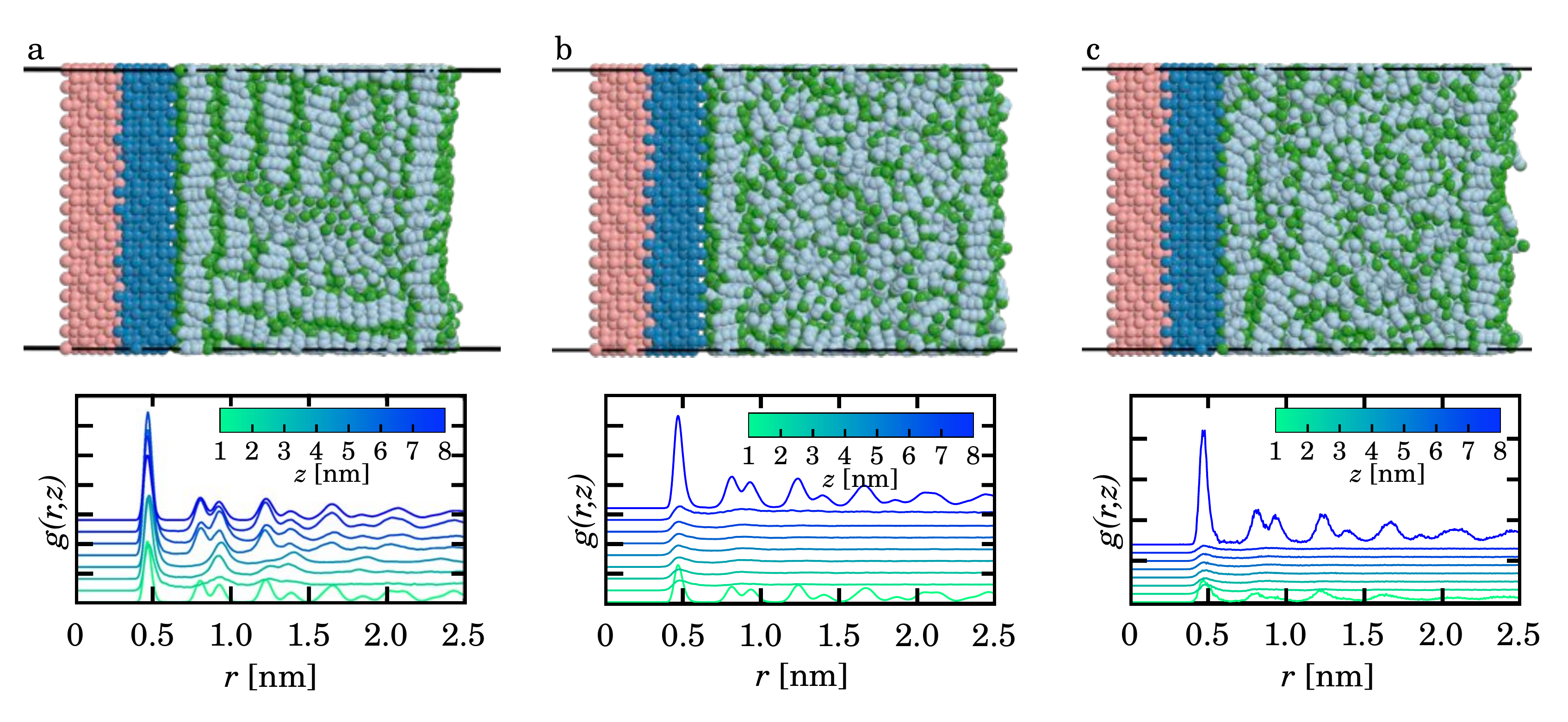}
		\caption{\label{fig:freezing}(a) Complete freezing of an $n$-octane film at 220~K, (b-c) Surface freezing of two $n$-octane films at 240~K in the vicinity of substrates with (b) $\epsilon_S=0.5$ and (c) $\epsilon_S=3.0$. Lateral radial distribution functions reveal long-range translation order across the film at 220~K, while the 240-K films are only translationally ordered at the solid-liquid and vapor-liquid interfaces.}
	\end{center}
\end{figure*}

\subsection{Qualitative Behavior of Octane Films\label{section:results:qualitative}}

As mentioned in Section~\ref{section:intro}, sufficiently long $n$-alkanes can undergo a process known as surface freezing~\cite{WuScience1993, OckoPRE1997}. To be more precise, alkane films behave differently at temperatures above and below $\widetilde{T}_m$, their effective  melting temperature. For films that are sufficiently thick $\widetilde{T}_m\approx T_m$, the equilibrium bulk melting temperature. At $T\le \widetilde{T}_m$, the entire film freezes into a lamellar crystal. At $ \widetilde{T}_m\le T\le T_s$, however, only a frozen monolayer emerges at the vapor-liquid interface. Here $T_s$ stands for the surface freezing temperature. Experimentally, surface freezing occurs for  $15\le n\le50$~\cite{TakiueJPCB2014}. Yet, we observe surface freezing in $n$-octane films simulated using the NERD force-field, even though $n=8$ for $n$-octane. For the films considered in this work, complete freezing occurs for $T\le 230$~K irrespective of the $\epsilon_S$ value (Fig.~\ref{fig:freezing}a). The existence of long-range lateral order in these 'cold` films is clearly visible in the lateral radial distribution functions depicted in Fig.~\ref{fig:freezing}a. For $235~\text{K}\le T\le250~$K, however, only a frozen monolayer emerges at the vapor-liquid and solid-liquid interfaces, with the center of the film remaining amorphous (Figs.~\ref{fig:freezing}b-c). Here, we define freezing in a purely structural sense, i.e.,~based on the presence of long-range lateral translational order. As will be further elucidated in Section~\ref{section:kinetics}, such frozen regions also correspond to dynamically decelerated regions in which translational and rotational degrees of freedom are decoupled. 

 The crossover temperature of $ \widetilde{T}_m=232.5\pm2.5$~K, is reasonably close to 216~K, the experimental melting temperature of $n$-octane~\cite{CRCHandbook}. (No computational estimate of $T_m$ is available for the NERD force-field.)  Our findings are also qualitatively consistent with earlier computational studies of $n$-octane~\cite{BitsanisJCP1996, DomnguezJCP2013}, and other closely-related $n$-alkanes~\cite{YamamotoJCP2000} using different force-fields, all reporting surface freezing.

\begin{figure}
	\begin{center}
		\includegraphics[width=0.3966\textwidth]{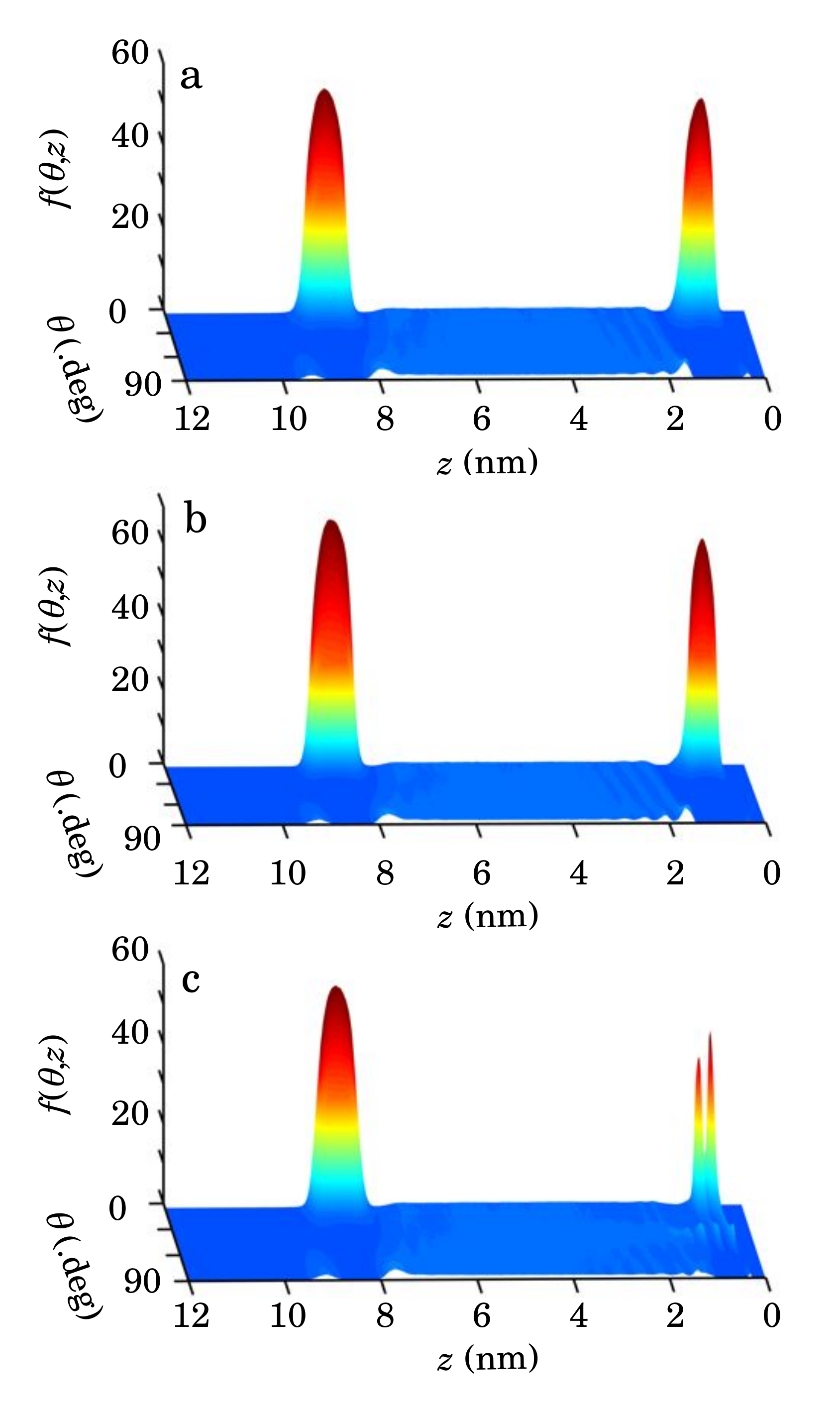}
		\caption{\label{fig:theta} Orientational distribution functions for a film at $T=240$~K and (a) $\epsilon_S=0.5$, (b) $\epsilon_S=1.0$,  (c) $\epsilon_S=3.0$. }
	\end{center}
\end{figure}

In order to examine the microstructure of  frozen monolayers, we compute the ODFs defined in Eq.~(\ref{eq:ODF}).   Fig.~\ref{fig:theta} depicts the ODFs for films at $T=240$~K. The frozen monolayers emerging at  vapor-liquid interfaces are characterized by a single peak in $f(\theta,z)$, corresponding to a perpendicular arrangement of octane molecules. Such an arrangement is  easily visible in the films depicted in Fig.~\ref{fig:freezing}a-c. In the vicinity of the substrate, however, the microstructure of the frozen monolayer depends  on $\epsilon_S$. For loose substrates, i.e.~$\epsilon_S=0.5$ (Fig.~\ref{fig:theta}a) and $\epsilon_S=1$ (Fig.~\ref{fig:theta}b), the microstructure closely resembles that of the free interface, as evident in the single peak of $f(\theta,z)$ at $\theta=0$ and $z\approx1$~nm. In contrast, a sticky substrate induces a different type of ordering, with the arising ODF having two distinct strong peaks at $\theta=0$: one at $z=0.65$~nm, and one at $z=0.9$~nm. The loci of these peaks are identical to the peaks in the molecular density profile depicted in the rightmost panel of Fig.~\ref{fig:density}. The existence of two peaks in $f(\theta,z)$ is a consequence of the corrugated surface of the 001 facet of fcc substrate. In other words, the molecules at the frozen monolayer can be present both in the valleys and peaks of the rough 001 surface. This changes their $z$ value even though they all have the same orientation with $\theta=0$.

\begin{figure}
\centering
\includegraphics[width=.45\textwidth]{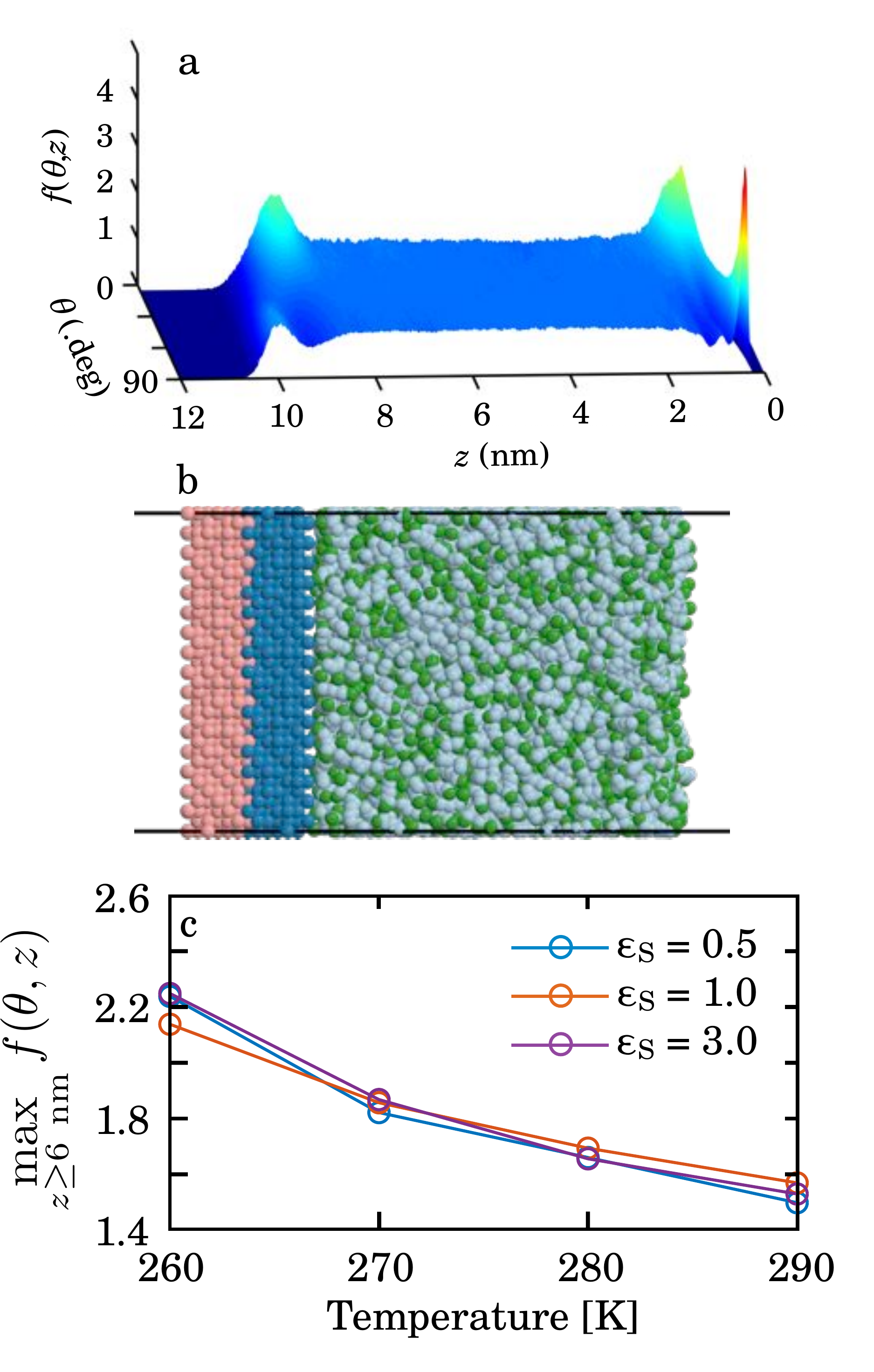}
\caption{\label{fig:pre-freezing}(a) Orientational distribution function at $T=270$~K and $\epsilon_S=0.5$, (b) A characteristic snapshot of the corresponding octane film. No freezing is observed. (c) Free surface alignment propensities at high temperatures.
}
\end{figure}

At higher temperatures, no frozen monolayer emerges at  vapor-liquid interfaces (Fig.~\ref{fig:pre-freezing}b). However,  $n$-octane molecules tend to have a mild preponderance to align along the $z$ axis, as evident in the ODF depicted in Fig.~\ref{fig:pre-freezing}a. Such a propensity can be qualitatively described as 'pre-freezing`, a phenomenon previously observed in earlier computational studies of alkane films~\cite{DomnguezJCP2013}. The notion of pre-freezing here is different from and should not be confused with the pre-freezing that is occasionally used for describing the interfacial freezing that precedes bulk freezing at $T>T_m$~\cite{KernJColloidIntSci1977}. We can quantify the extent for such pre-freezing with the following two order parameters. The first one is $\xi=\max_{z\ge~6~\text{nm}}f(\theta,z)$,  the maximum of the free-surface peak of ODF. Note that $\xi=1$ for a fully isotropic film. Any deviation from unity will henceforth correspond to broken rotational symmetry. As depicted in Fig.~\ref{fig:pre-freezing}c, $\xi$ is always significantly larger than unity, even for the films at $T=290$~K. Also the peak always occurs at $\theta^*=\text{argmax}_{z\ge~6~\text{nm}}f(\theta,z)=0$ irrespective of $T$.   As expected, pre-freezing becomes stronger at lower temperatures.

\begin{figure*}
\centering
\includegraphics[width=.9\textwidth]{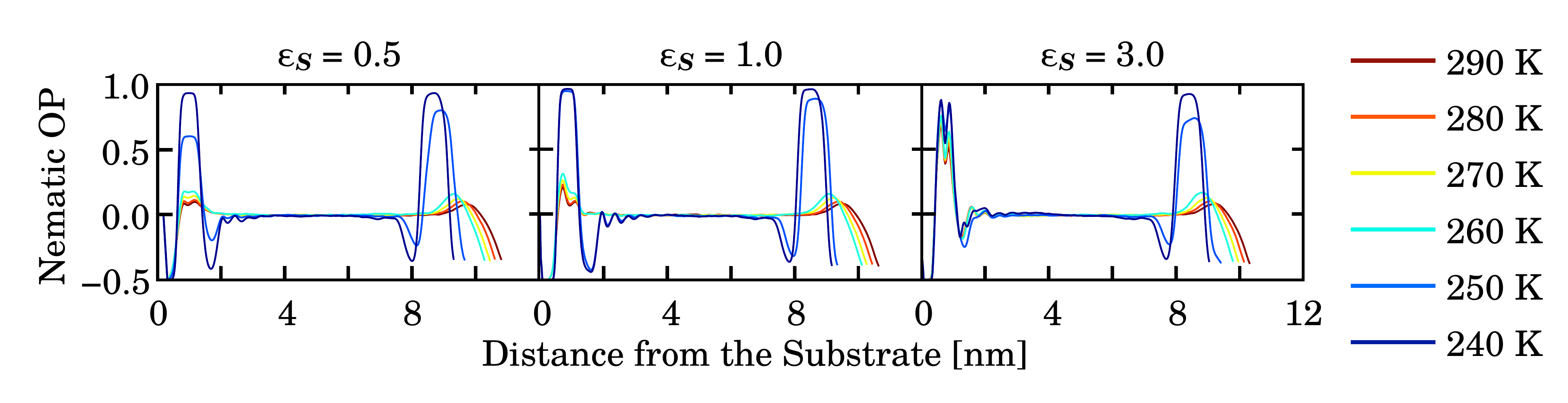}
\caption{\label{fig:theta-op}Profiles of nematic order parameter for different temperatures and $\epsilon_S$ values.}
\end{figure*}

The second order parameter that is adopted from the liquid crystal literature is called the \emph{nematic order parameter (OP)} and is essentially the second moment of the ODF~\cite{HajiAkbariJPhysA2015}:
\begin{eqnarray}
S_{zz}(z) &=& \frac12\int_0^{\frac{\pi}2}\left[3\cos^2\theta-1\right]f(\theta,z)\sin\theta d\theta
\end{eqnarray}
For a fully isotropic fluid, $S_{zz}=0$, while the values of $+1$ and $-0.5$ correspond to perfect alignment along and perpendicular to the $z$ axis, respectively. Fig.~\ref{fig:theta-op} depicts $S_{zz}(z)$ profiles  for different films. Note that the nematic OP is very close to unity for surface-frozen monolayers at low temperatures. Surface pre-freezing of high-temperature films is manifest in positive peaks of the nematic OP corresponding to weak alignment along the $z$ direction. Similar to $\xi$ that decreases with $T$, the heights of these peaks diminish as $T$ increases.

Similar to the free interface, a mildly pre-frozen monolayer emerges in the vicinity of loose substrates ($\epsilon_S=0.5, 1.0$), with characteristic mild peaks in the nematic OP profiles (Fig.~\ref{fig:theta-op}). For sticky substrates, however, a frozen monolayer emerges at all temperatures, with its ODF resembling the one depicted in Fig.~\ref{fig:theta}c.  The nematic order parameter also demonstrates two strong peaks close to unity, consistent with perfect alignment along the $z$ direction spatially modulated by the corrugated surface of the substrate.

\subsection{Thermodynamic Properties\label{section:thermo}}

\begin{figure*}
	\begin{center}
	\includegraphics[width=0.8509\textwidth]{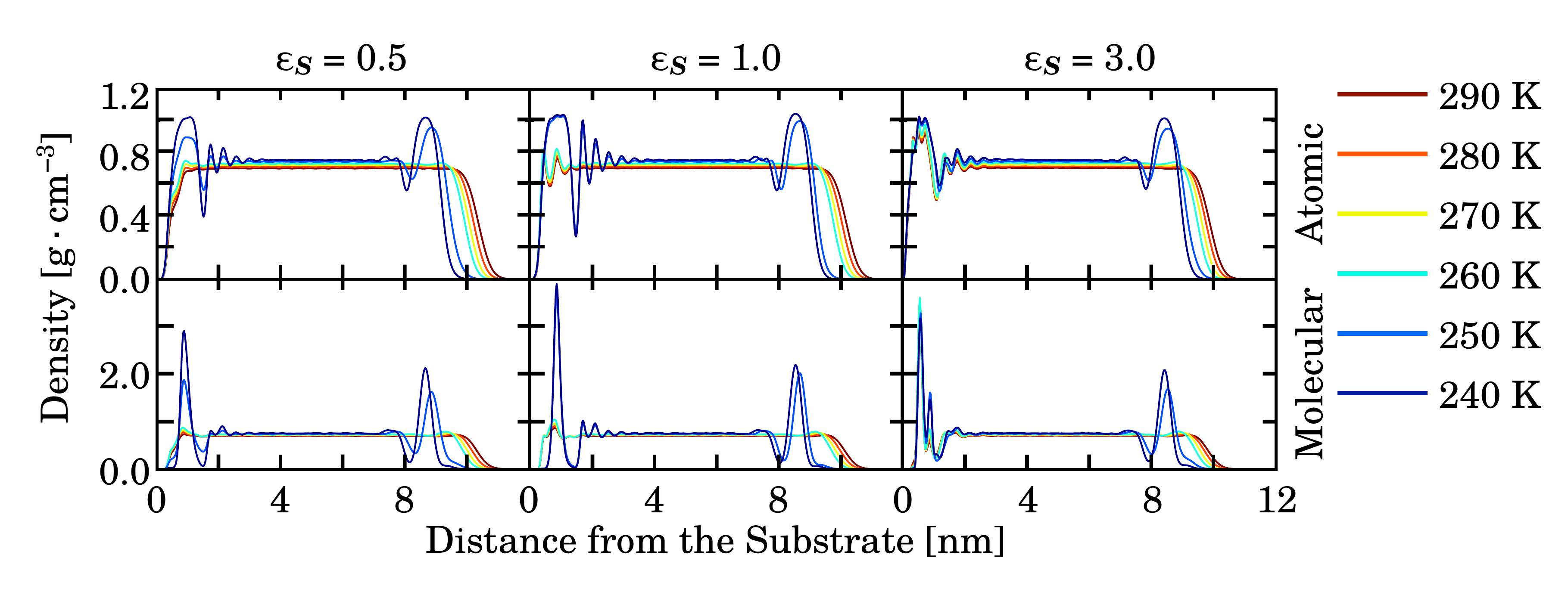}
	\caption{\label{fig:density}Profiles of atomic and molecular density for different temperatures and $\epsilon_S$ values.}
	\end{center}
\end{figure*}

\noindent\textbf{Density}:
Profiles of atomic and molecular densities are depicted in Fig.~\ref{fig:density}. Both  profiles are peaked at the frozen monolayers that emerge at the vapor-liquid and solid-liquid interfaces. In the case of free interfaces and loosely attractive substrates, a single large peak is observed. For sticky substrates, however, the peak splits into two as a result of spatial modulation induced by the corrugated substrate. The loci of these peaks correspond to the maxima of ODF (Fig.~\ref{fig:pre-freezing}) and nematic OP (Fig.~\ref{fig:theta-op}). Such large peaks are always followed by distinct valleys that correspond to the crystal/liquid interface. As evident in Fig.~\ref{fig:theta-op}, the nematic OP is negative at these valleys as the octane molecules have a (weak) propensity to align parallel to the substrate. 

\begin{figure}
\centering
\includegraphics[width=.45\textwidth]{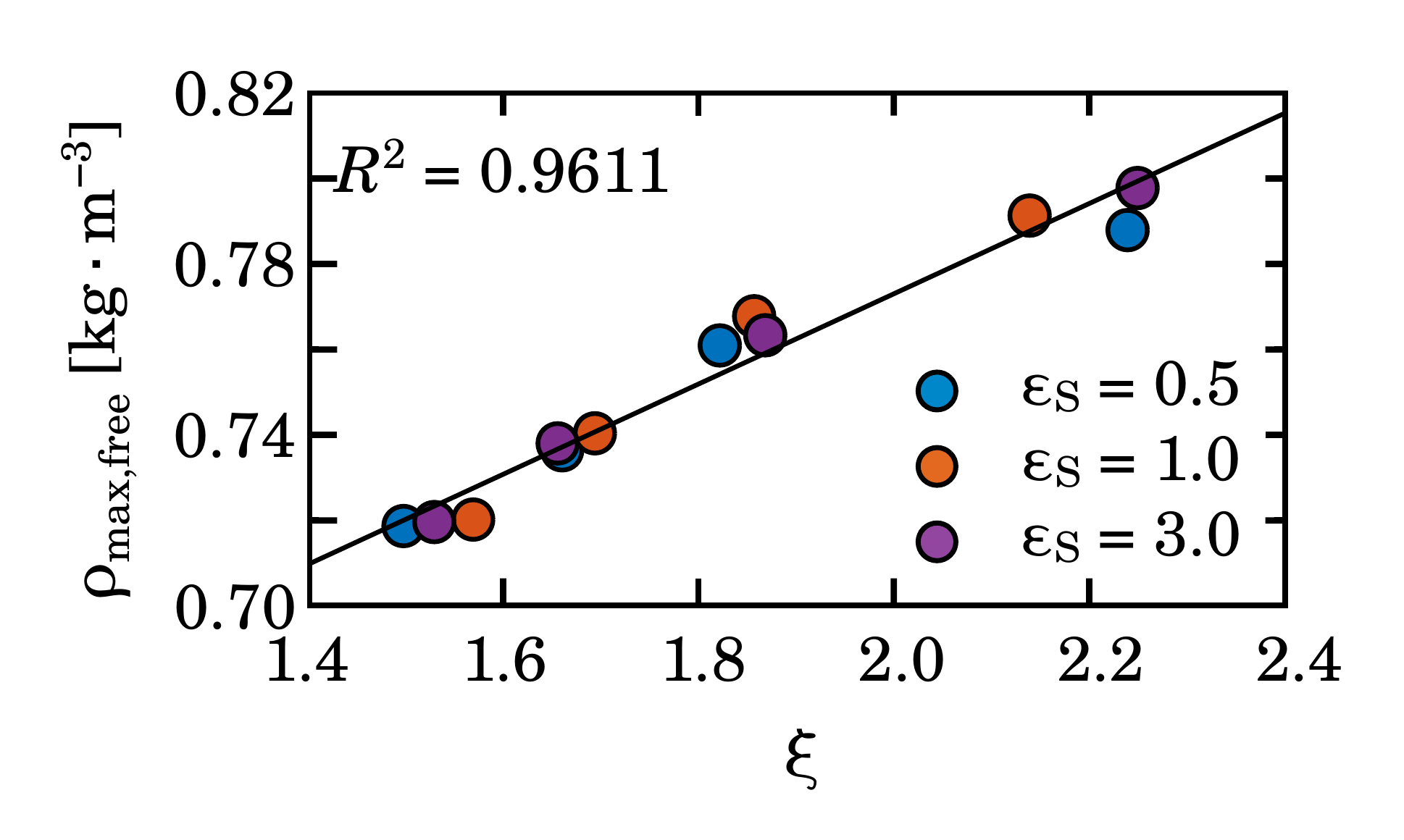}
\caption{\label{fig:density-vs-xi}Linear correlation between $\xi$ and $\rho_{\text{max,free}}$. The dark line is a linear fit to the individual data points.  }
\end{figure}

A very interesting feature of octane films is the stratification of octane molecules at free interfaces even in the absence of surface freezing, i.e.,~at $T>250$~K. This behavior is contrary to what is observed in atomic thin films in which density drops monotonically across a free interface~\cite{HajiAkbariJCP2014, ShiJCP2011}. A single mild peak emerges in the atomic and molecular density profile, with its amplitude closely correlating with the pre-freezing order parameters defined in Section~\ref{section:results:qualitative}. A linear correlation between $\xi$ (Fig.~\ref{fig:pre-freezing}c) and $\rho_{\text{max,free}}$, the amplitude of the molecular density peak, is depicted in Fig.~\ref{fig:density-vs-xi}. Henceforth, the nonmonotonic behavior of density is a result of pre-freezing that creates local high-density regions in the interfacial region. Density oscillations at a free interface have been previously observed in experimental and computational studies of not only  alkanes~\cite{HarrisJPC1992, DomnguezJCP2013} but also alkali metals~\cite{RiceJNonCrystSol1996, RiceJCP1998} and ionic liquids~\cite{JiangJPCC2008}. The emergence of density oscillations has been attributed to a wide range of factors, including large separation between the critical and triple-point temperatures~\cite{VelascoJCP2002} and building block anisotropy~\cite{MederosPRA1992}. %Such factors can overcome the ability of capillary waves to dampen such density oscillations~\cite{TarazonaJCP2002}. 
Unlike the atomic thin films, the oscillatory density profiles in octane films cannot be fitted to the commonly used hyperbolic tangent functional form~\cite{SidesPRE1999}, henceforth making the determination of the width of the free interface based on density nontrivial.

Close to the substrates, both  atomic and molecular density profiles are oscillatory. This is consistent with the traditional picture of confinement in which a wall induces structure in the surrounding liquid~\cite{PlischkeJCP1986, DickmanJCP1988, CurroJCP1994}. The stratification of the octane liquid is, however, an interplay between the  structuring induced by the substrate and the natural propensity of the interfacial region to pre-freeze. Density oscillations in the vicinity of a loosely attractive substrate ($\epsilon_S=0.5$) are very similar to the profiles at the free interface, with the substrate inducing very little structure in the liquid. This is very similar to theoretical predictions obtained from density functional theory for associating fluids near a hard wall~\cite{CurroJCP1994}.  As $\epsilon_S$ increases, however, stratification becomes more pronounced.  For instance, multiple peaks in density emerge at $\epsilon_S=1.0$, even though the solid-liquid interface has the same orientational fingerprints as the vapor-liquid interface.

\begin{figure*}
	\begin{center}
	\includegraphics[width=0.8450\textwidth]{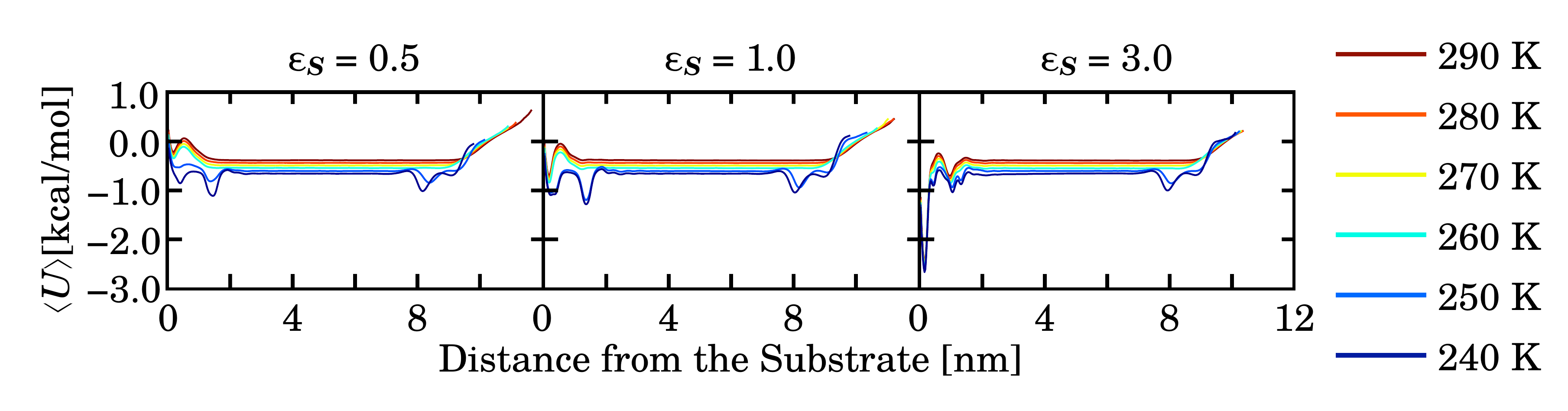}
	\caption{\label{fig:energy}Profiles of potential energy for different temperatures and $\epsilon_S$ values.}
	\end{center}
\end{figure*}

\noindent\textbf{Potential Energy and Internal Structure}:
Profiles of $\langle U\rangle$, the average potential energy, are depicted in Fig.~\ref{fig:energy}. $\langle U\rangle$ exhibits a maximum at the center of a frozen monolayer. This apparent energetic penalty can be attributed to the lamellar microstructure of a monolayer in which the united-atom CH$_2$ sites cluster in the middle. Since $|\epsilon_{\text{CH}_2,\text{CH}_2}|<|\epsilon_{\text{CH}_2,\text{CH}_3}|<|\epsilon_{\text{CH}_3,\text{CH}_3}|$, such CH$_2$-rich regions will correspond to higher potential energies. This energetic penalty is, however, offset by the CH$_3$-rich regions at  two ends of a monolayer, corresponding to minima in $\langle U\rangle$.  It is also noteworthy that  $z^*$, the locus of the maximum of $\langle U\rangle$, matches the corresponding $z^*$ for density, $f(\theta,z)$ and $S_{zz}(z)$, further confirming that all these oscillations are manifestations of the same phenomenon, i.e.,~interfacial freezing. Like density, $\langle U\rangle$ profiles are oscillatory close to the substrate, even in the absence of interfacial freezing, with the amplitudes and number of  oscillations increasing with $1/T$ and $\epsilon_S$. However, we do not observe any nonmonotonicity in the pre-frozen free interfaces, suggesting that the propensity of a surface to pre-freeze has a purely entropic origin. 

\begin{figure*}
	\begin{center}
	\includegraphics[width=0.8450\textwidth]{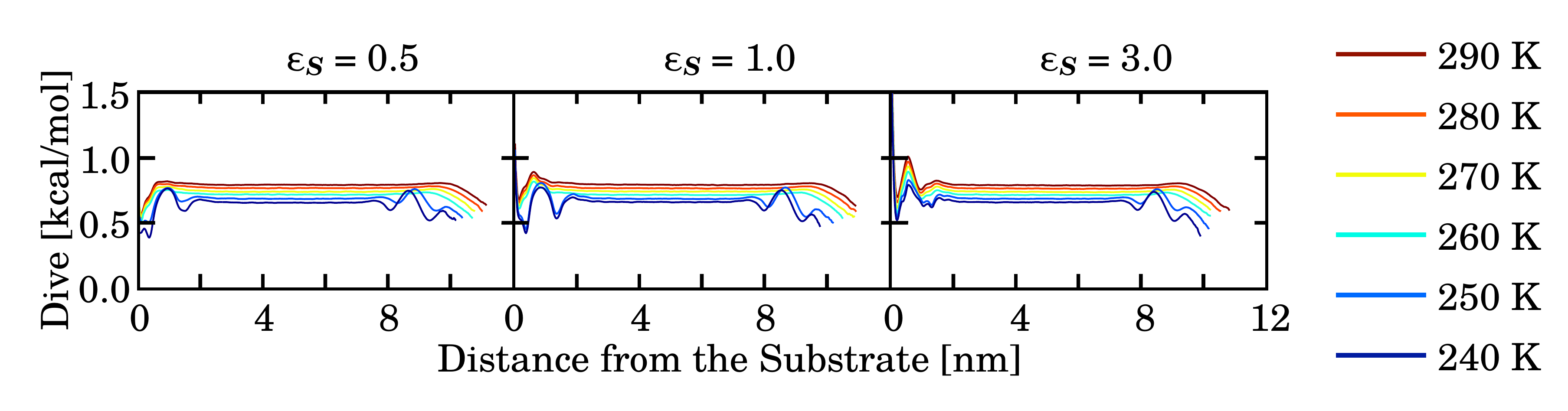}
	\caption{\label{fig:dive}Profiles of dive energy  for different temperatures and $\epsilon_S$ values.}
	\end{center}
\end{figure*}

Although potential energy profiles provide valuable information about the relative energetic stability of different regions of a confined material, they convey very little about the ability of a material to explore its confinement-induced potential energy landscape. A systematic way of characterizing the latter is to obtain an ensemble of inherent structures, and to compute the amount of decrease in potential energy for different regions of the material~\cite{ShiJCP2011}. Here, $z$ binning is performed based on the initial (pre-minimization) positions of the individual atoms. Spatial profiles of this quantity, which we denoted as \emph{dive} profiles in our earlier publication~\cite{HajiAkbariJCP2014}, are depicted in Fig.~\ref{fig:dive}. For all the films, whether they go through surface freezing or pre-freezing, no noticeable overall difference is detected between the landscape depth in the bulk and at the free interface. The ability of a free interface to modify the potential energy landscape accessible to a liquid is, therefore, far more limited in the case of $n$-octane. This is unlike the behavior observed in the atomic thin films considered in Refs.~\cite{ShiJCP2011, HajiAkbariJCP2014} in which a vapor-liquid interface increases the depth of the potential energy landscape accessible to the liquid at its vicinity. Such a heightened access to the minima of the potential energy landscape appears to be key in the ability of a material to form ultrastable glasses upon vapor deposition~\cite{RezaJPhysCondMatt2015}. By this token, the ability of $n$-alkanes to form ultrastable glasses will be limited considering the lack of elevated access to the potential energy landscape of the liquid at the free interface.

\begin{figure*}
	\begin{center}
	\includegraphics[width=0.6617\textwidth]{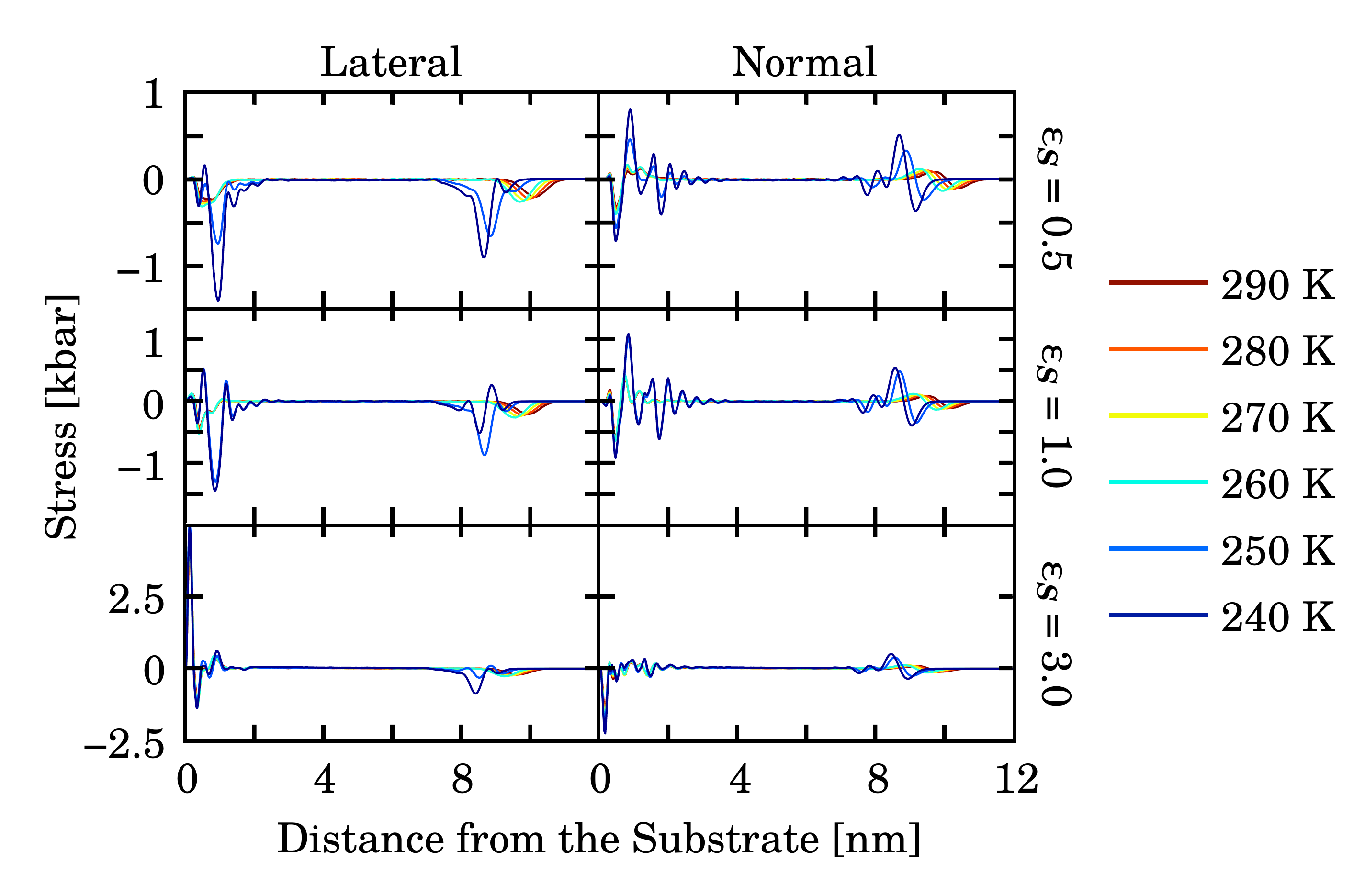}
	\caption{\label{fig:stress}Profiles of normal and lateral stress for different temperatures and $\epsilon_S$ values.}
	\end{center}
\end{figure*}

\noindent\textbf{Stress}:
Of all the thermodynamic quantities considered in this work, none oscillates more strongly than  normal and lateral stress  (Fig.~\ref{fig:stress}). Such oscillations are particularly strong in frozen monolayers at both interfaces, and extend well beyond the frozen region. Indeed, sharp interfaces are maintained as a result of surface tension that stems directly from anisotropies of the stress tensor~\cite{GloorJCP2005}. In the case of frozen monolayers, in particular, two such sharp interfaces are present: one between the substrate (or vapor) and the frozen monolayer, and the other between the monolayer and the liquid.  Consequently, oscillations in stress occur across a relatively wider region of the film. 

It can be argued that stress anisotropy is among the most systematic ways of defining what constitutes an interfacial region. Indeed, a region with  anisotropic stress responds differently to mechanical stress, and will thus have mechanical properties distinct from the bulk. In this work, we define the bulk region as the largest connected region in which the difference between normal and lateral stress does not exceed 20~bar. Choosing this cutoff, which is always less than 10 per cent of the largest difference between lateral and normal stress at $T=$290~K, is necessary considering the difficulty of converging virial-based estimates of stress in molecular simulations. Applying this criterion yields an interfacial width that increases with temperature, from $\approx$2.30~nm at 260~K to $\approx$2.45~nm at 290~K. The interfacial width, as determined from the anisotropy in stress tensor, is very difficult to measure in experiments. Nevertheless, the temperature scaling of the dynamical length scales in thin films of polystyrene glasses reveals a similar dependence of temperature, with the width of the highly mobile region increasing with temperature~\cite{PaengJACS2011}. It can indeed be argued that a correlation must exist between the thermodynamic length scale (obtained from stress tensor) and the dynamical length scales (defined as per mobility).

\begin{figure*}
	\begin{center}
	\includegraphics[width=0.7825\textwidth]{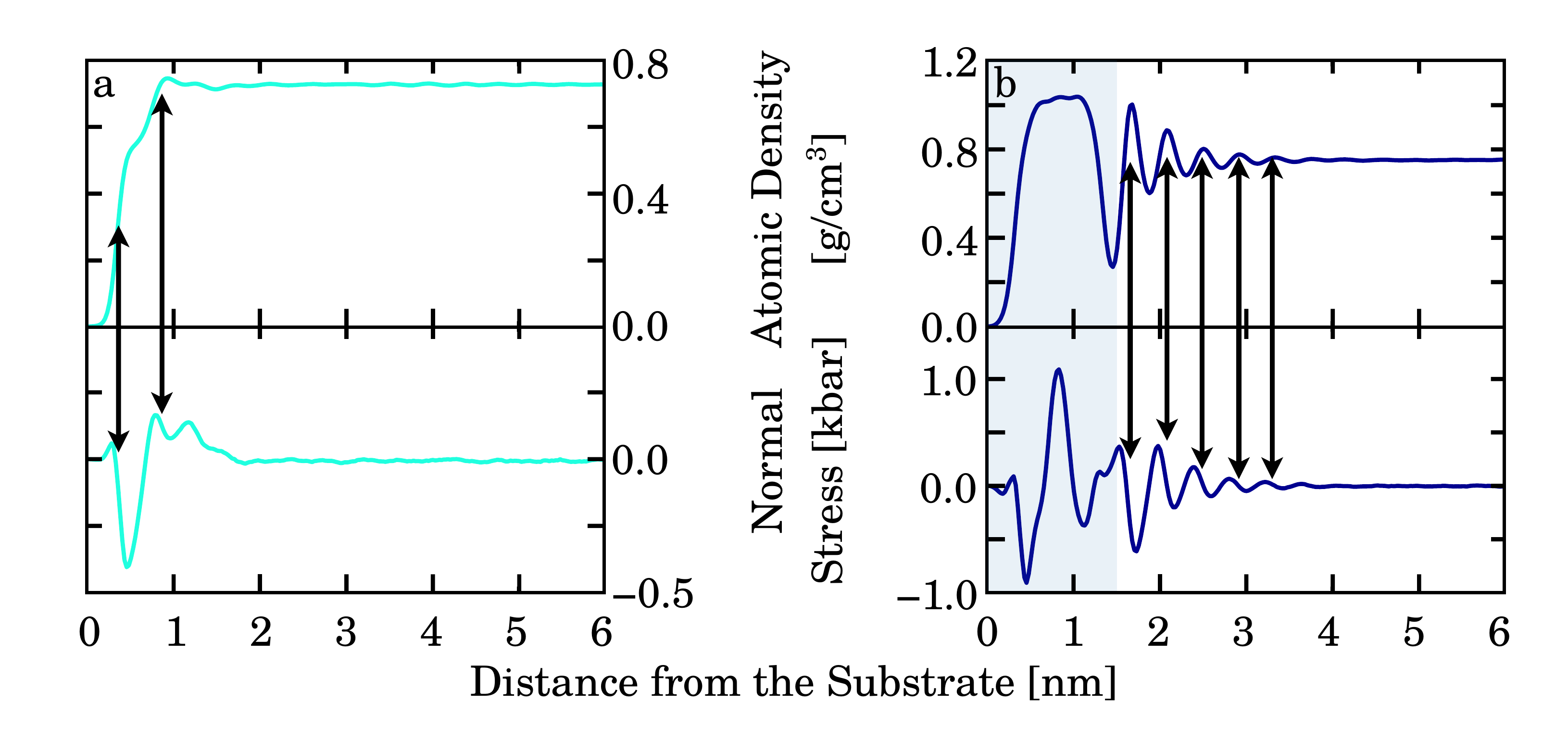}
	\caption{\label{fig:normal-vs-density}Correlation between the profiles of atomic density and normal stress for a film at (a) $T=260$~K, $\epsilon_S=0.5$, and (b) $T=240$~K, $\epsilon_S=1.0$. The region shaded in light blue has long-range translational order.}
	\end{center}
\end{figure*}

\begin{figure*}
	\begin{center}
	\includegraphics[width=.9\textwidth]{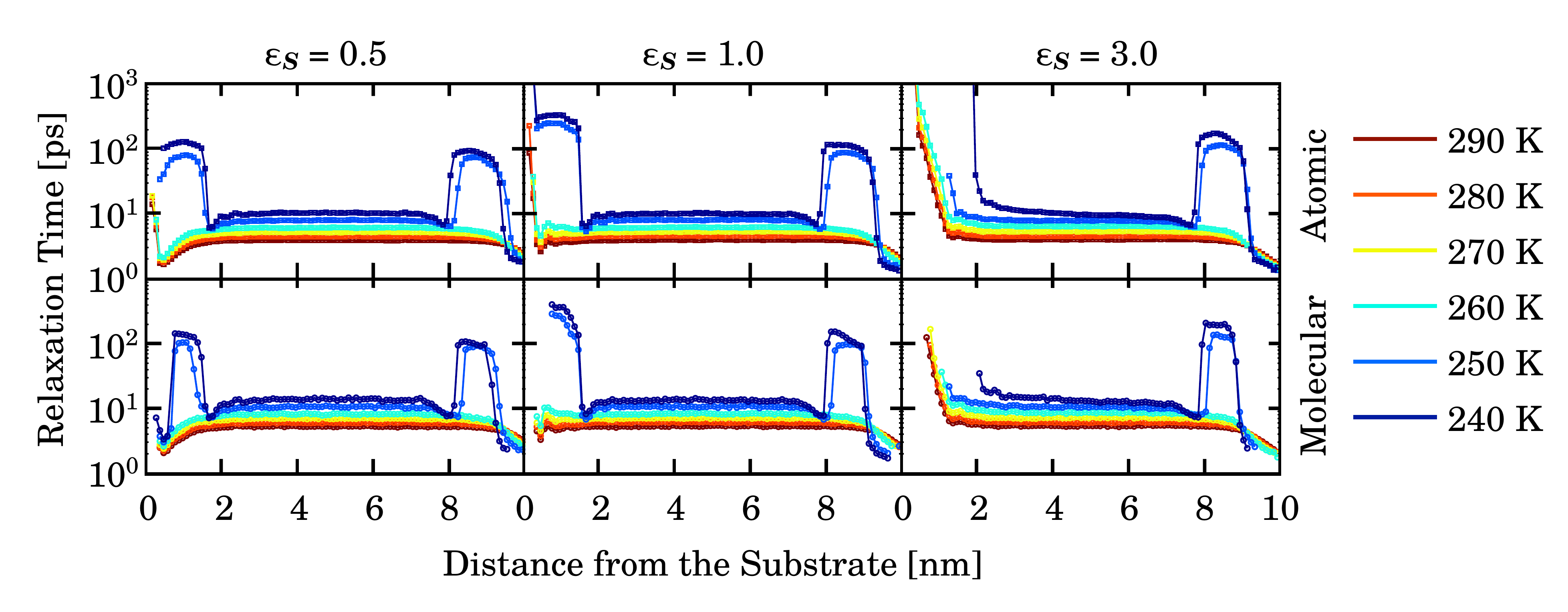}
	\caption{\label{fig:trans}Profiles of atomic and molecular translational relaxation time for different temperatures and $\epsilon_S$ values.}
	\end{center}
\end{figure*}

\begin{figure*}
	\begin{center}
	\includegraphics[width=.9\textwidth]{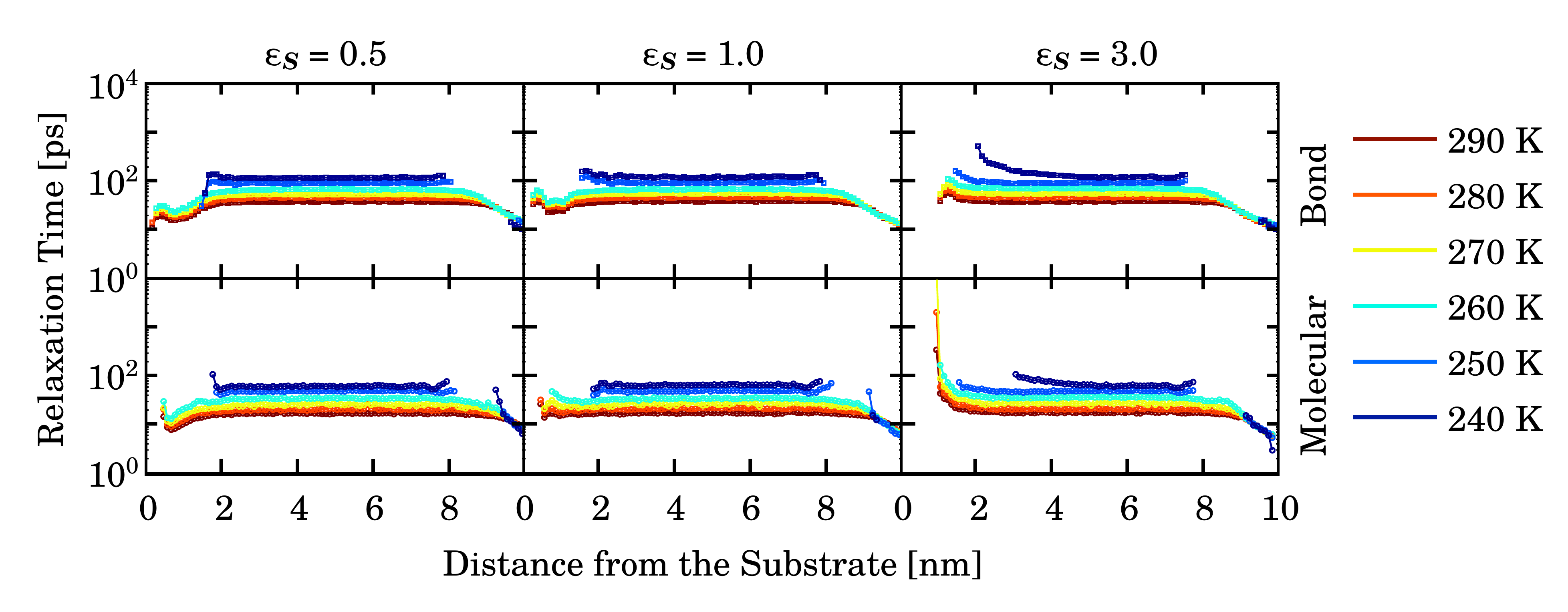}
	\caption{\label{fig:rot}Profiles of bond and molecular rotational relaxation time for different temperatures and $\epsilon_S$ values.}
	\end{center}
\end{figure*}

In our earlier work~\cite{HajiAkbariJCP2014}, we established a strong correlation between the substrate-induced profiles of normal density and normal stress in amorphous regions of atomic thin films. By inspecting the density and normal stress profiles, we explore the existence of such a correlation in octane films. Here, the quantity of relevance is the atomic-- and not the molecular-- density, as the positions of individual molecules are not sufficient for determining the stress tensor. We observe a correlation between atomic density and normal stress in amorphous regions of octane films. For loosely attractive substrates (Fig.~\ref{fig:normal-vs-density}a), the correlation is weaker to the extent that the first maximum in normal stress only corresponds to a shoulder in atomic density. In other words, the solid-liquid interfacial region is very similar to a free interface, as the substrate exerts minimal effect in ordering and stratifying the liquid in its vicinity. In the case of stickier substrates  (Fig.~\ref{fig:normal-vs-density}b), however, the correlation is much more pronounced, and the peaks and valleys of atomic density and normal stress follow one another more closely. It is, of course, necessary to emphasize the absence of such correlations in parts of the film that have long-range translational order, as crystals can respond to stress in nontrivial ways.

\begin{figure}
	\begin{center}
	\includegraphics[width=0.45\textwidth]{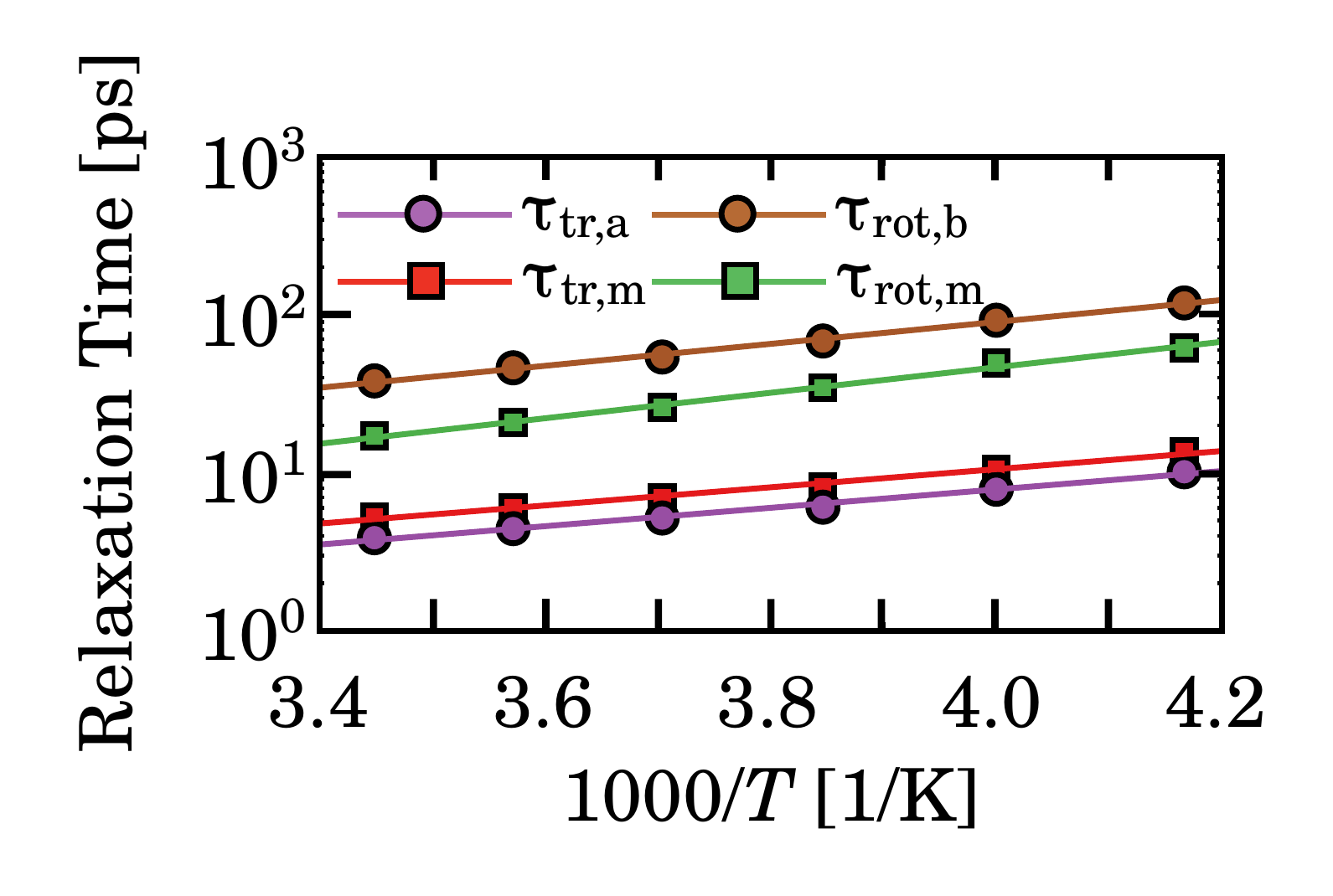}
	\caption{\label{fig:bulk-rlx-r5}Temperature dependence of bulk relaxation times for $\epsilon_S=0.5$.}
	\end{center}
\end{figure}

\begin{figure*}
	\begin{center}
	\includegraphics[width=0.7825\textwidth]{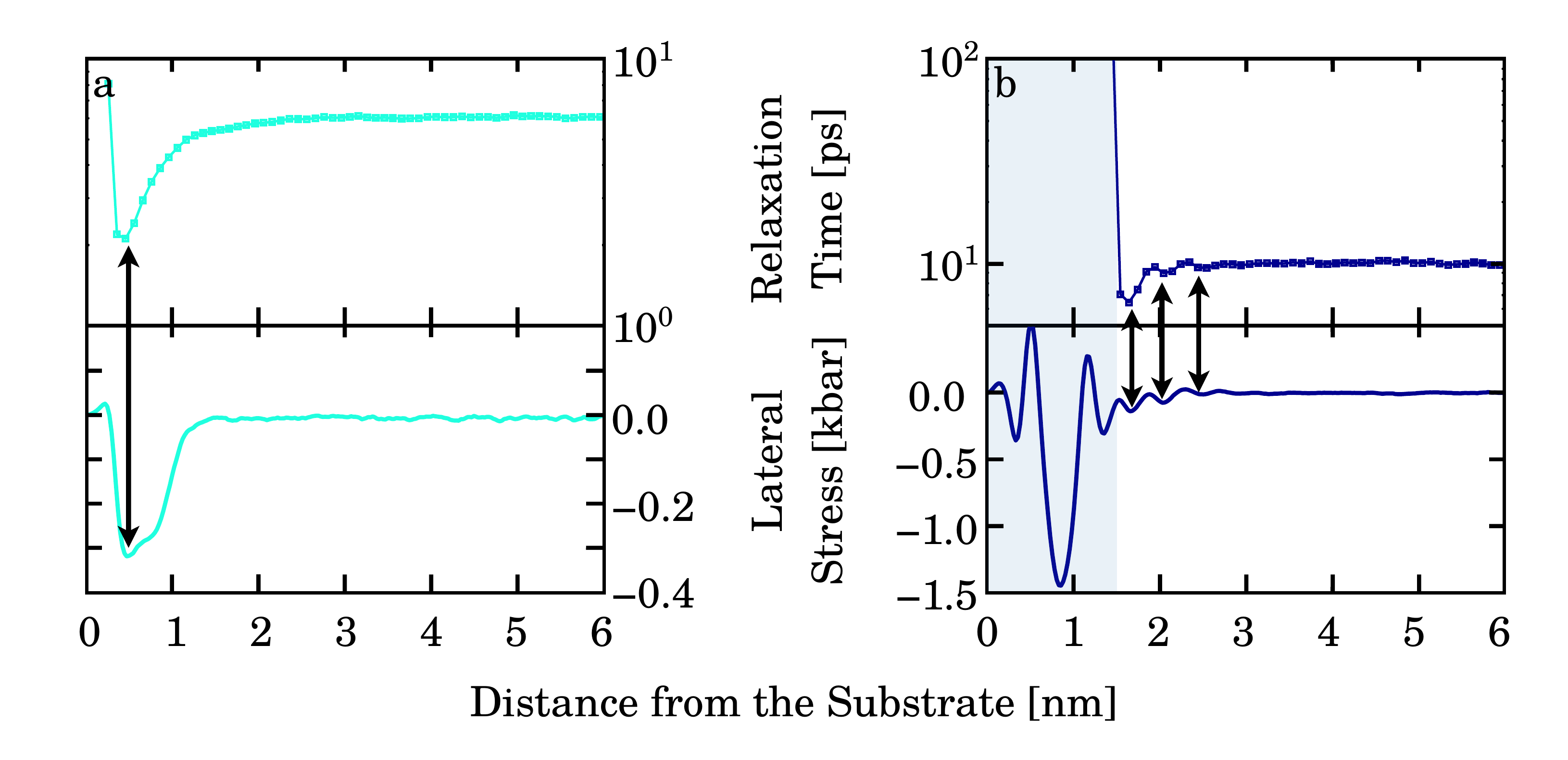}
	\caption{\label{fig:lateral-vs-trans}Correlation between the profiles of atomic translational relaxation time and lateral stress for a film at (a) $T=260$~K, $\epsilon_S=0.5$, and (b) $T=240$~K, $\epsilon_S=1.0$. The region shaded in light blue has long-range translational order.}
	\end{center}
\end{figure*}

\subsection{Kinetic Properties\label{section:kinetics}}

Figs.~\ref{fig:trans} and~\ref{fig:rot} depict  profiles of translational and rotational relaxation times. In  surface-frozen monolayers, a sharp increase in relaxation times is observed, with both the atomic and molecular relaxations times increasing by at least one order of magnitude with respect to their bulk values. We observe an even larger increase in molecular and bond orientational relaxation times. We are unable to quantify the extent of such an increase as the  orientational autocorrelation functions given in Eqs.~(\ref{eq:hm}) and (\ref{eq:hb}) never relax to zero in the timescale of our simulations.  This decoupling of translational and orientational degrees of freedom is a direct consequence of layering at the surface, and is common in phases with long-range orientational order such as liquid crystals~\cite{LeeJCP1987}. In such systems, the relaxation of translational degrees of freedom proceeds through a slipping mechanism that does not require orientational relaxation.

In Section~\ref{section:thermo}, we discuss non-monotonicities in several thermodynamic quantities, including $\xi$, $S_{zz}$ and density, across free interfaces of the films that do not undergo surface freezing. We do not detect any such non-monotonicity in relaxation time profiles. Yet, we are  unable to rule out its possibility for the following reasons. First of all, the uncertainties associated with computing dynamical quantities such as relaxation times are generally larger than the uncertainties in estimating thermodynamic quantities such as density. After all, the former are computed from auto-correlation functions while the later are obtained from simple time averaging. Secondly, the convention used  for defining $z$-dependent autocorrelation functions will inevitably lead to some mixing between neighboring slices as the molecules that  contribute to the autocorrelation function of a particular $z$ slice can leave and re-enter that slice. Such a procedure can mask the existence of subtle spatial differences in relaxation time, a situation that is completely plausible in the pre-frozen interfaces considered in this work. 

The dynamical features of a film in the vicinity of a substrate depend on $\epsilon_S$. Like the free interface, the surface-frozen monolayers close to loose substrates demonstrate a deceleration of dynamics and a decoupling of translational and rotational degrees of freedom. This slow-down happens in the vicinity of a sticky substrate as well. However, it is not clear whether translational and rotational degrees of freedom decouple as the corresponding relaxation times cannot be computed in the timescale of our simulations. For the films that do not undergo surface freezing, however, two distinct dynamical regimes are observed. For sticky walls ($\epsilon_S=3.0$), the dynamics is decelerated close to the substrate, even in amorphous regions of the film. Indeed,  relaxation times can be as much as two orders of magnitude larger than the corresponding bulk values. In the vicinity of loose substrates ($\epsilon_S=0.5$), however, the dynamics is accelerated, as in atomic thin films. For  $\epsilon_S=1.0$, the relaxation time profiles are more oscillatory with the overall dynamics becoming only slightly faster with respect to the bulk. This is consistent with the behavior observed in atomic thin films that also demonstrate two distinct dynamical regimes in the vicinity of sticky and loose substrates. 

Fig.~\ref{fig:bulk-rlx-r5} depicts the temperature dependence of bulk relaxation times for $\epsilon_S=0.5$. Those are obtained from averaging the relaxation time profiles of Figs.~\ref{fig:trans} and \ref{fig:rot} over $z[\text{nm}]\in [2.5,7.5]$. An identical behavior is observed for other $\epsilon_S$ values, with the results not shown for conciseness.  All relaxation times demonstrate an Arrhneius-type dependence on temperature, with no sign of fragility. This is not unusual considering the earlier  viscosity measurements for $n$-octane~\cite{LewisJCP1965, MillerJPolymSci1968}, which predict a glass transition temperature of $T_g=85$~K, markedly lower than the temperatures considered here. 
% Even though there are liquids that are fragile at temperatures way above $T_g$~\cite{LaughlinJPhysChem1972}, most liquids demonstrate fragility only at the deeply supercooled regime, and close to the glass transition temperature.
%Indeed, short-chain $n$-alkanes are known to be poor glass-formers due to their high propensity to crystallization~\cite{TabataChemPhysLett1976}. 
Note that atomic translational relaxation times are always lower than their molecular counterparts, due to the relative ease of relaxing atomic degrees of freedom. It is, however, not possible to systematically compare the bond and molecular orientational relaxation times, since they correspond to relaxation features of different Legendre polynomials. Therefore, the fact the bond relaxation times are higher does not have any physical meaning. Similarly, it is not meaningful to compare translational and orientational relaxation times, as they have also been computed from decays of different autocorrelation functions.

In our earlier work~\cite{HajiAkbariJCP2014}, we established a close correlation between lateral relaxation time and lateral stress profiles in amorphous regions of atomic thin films. This is consistent with what happens in simple liquids in which diffusivity increases upon decreasing pressure~\cite{MukherjeeJCP2002}. We find a similar correlation between the profiles of atomic translational relaxation time and lateral stress in octane films.  Fig.~\ref{fig:lateral-vs-trans} shows such a correlation for two representative films, with the peaks and valleys of $\tau_{\text{tr,a}}$ and lateral stress closely following one another in  the amorphous region (determined from RDF). It is interesting to note that such a correlation can also be rationalized by the fact that the compressibility of a simple fluid increases upon decreasing pressure, and this enhances structural relaxation by facilitating local density fluctuations.

\section{Conclusions\label{section:conclusions}}

In this work, we study the effect of substrate on thermodynamic and kinetic anisotropies in $n$-octane thin films. We observe complete freezing at temperatures below $232.5\pm2.5$~K. For $235~\text{K}\le T\le250~\text{K}$, however, a frozen monolayer emerges at vapor-liquid interfaces, as well as in the vicinity of loosely attractive substrates. Such frozen monolayers correspond to large peaks in several thermodynamic and kinetic properties, such as density, potential energy, nematic order parameter, and translational and rotational relaxation times. Also, translational and rotational degrees of freedom are decoupled in the monolayer due to long-range orientational ordering of the $n$-octane molecules. 

At higher temperatures, interfacial freezing only occurs in the vicinity of sticky substrates. At vapor-liquid interfaces, and close to loose substrates, only a weak propensity is observed for the molecules to 'pre-freeze` and align perpendicular to the interface. Such a propensity manifest itself in mild peaks in atomic and molecular density and $f(\theta,z)$ and $S_{zz}$ orientational order parameters.

The amorphous regions of the films are stratified in the vicinity of  substrates, with the extent of stratification increasing upon decreasing temperature, or increasing the interaction parameter $\epsilon_S$. Density oscillations are minimal close to loose substrates, and the interface almost resembles a vapor-liquid interface. Similar to density, oscillations are observed in other thermodynamic quantities such as potential energy, dive (as a result of energy minimization), stress, and orientational order parameters. In the vicinity of a substrate, we  confirm the existence of correlation between atomic density and normal stress in amorphous regions of the film, consistent with our observations in atomic thin films.  We also observe two distinct dynamical regimes at the solid-liquid interface. In the vicinity of loose substrates, dynamics is accelerated, while a sticky substrate decelerates dynamics in its vicinity. Finally, we are able to establish a correlation between lateral atomic translational relaxation times and lateral stress in amorphous regions of the films. 

Short-chain $n$-alkanes are known to be poor glass formers due to their strong propensity to crystallize~\cite{TabataChemPhysLett1976}. Our findings suggest that the vapor deposition process that yields ultrastable glasses for many materials is unlikely to be very successful in the case of $n$-alkanes. This is not only due to the propensity of alkane films to undergo surface freezing, but also because of the fact that the free interface does not create a more accessible potential energy landscape even when the alkane film is amorphous.  It has, indeed, been recently demonstrated that there are materials that cannot form a ultrastable glass upon vapor deposition~\cite{WubbenhorstNatComm2012}, unlike  earlier suggestions that this process might be universal~\cite{ZhuCPL2010}. This has led to speculations that a correlation might exist between the fragility of a liquid and its ability to form an ultrastable glass~\cite{AngellJNonCrystSol2015}. This correlation is not clear-cut as some relatively strong liquids still form ultrastable glasses upon vapor deposition~\cite{SepulvedaPRL2014}. There might, however, be other dimensions to this problem, namely the presence of a mobile interfacial region, as well the accessibility of deeper minima of the potential energy landscape at the interface. If this picture is accurate, $n$-octane must not yield ultrastable glasses upon physical vapor deposition. It is therefore a worthwhile experimental endeavor to investigate vapor-deposited glasses of $n$-alkanes, and other molecules with aliphatic chains to determine whether this picture is accurate.  

It is necessary to emphasize that, according to this picture, elevated orientational ordering at a free interface will not necessarily make a material a poor ultrastable glass former. Indeed, experimental~\cite{DalalJPCB2013, ShakeelPNAS2015, AnkitChemMater2015} and computational~\cite{LyubimovJCP2015} studies of ultrastable glasses have confirmed the existence of such ordering at vapor-liquid interfaces of good ultrastable glass formers.

{It is necessary to emphasize that the microstructure of the liquid in the vicinity of the solid-liquid interface might depend on the structure of the substrate. For instance, the orientational order that would emerge close to a sticky structureless substrate will be different from what is observed here, as reported in Ref.~\cite{YamamotoJCP2007}. Since the liquid that is close to a low-$\epsilon_S$ substrate is structurally very similar to the liquid at the free interface, we expect that such differences will only be important at large values of $\epsilon_S$.
%We, however, think that such differences will only be important at larger values of $\epsilon_S$ for the following reason. As demonstrated in Section~\ref{section:results:qualitative}, the liquid that is close to a low-$\epsilon_S$ wall essentially looks the same as the liquid at the free interface. 
Even then, the observed deceleration of the dynamics (with respect to bulk) is not expected to disappear if a structureless sticky substrate is utilized. It is, however, interesting to study the combined effect of $\epsilon_S$ and substrate corrugation on the microstructure of the frozen monolayers, as well as the thermodynamic and kinetic anisotropies in the film.
}

There are other interesting questions to be asked about alkane films  besides their ability or lack thereof to form ultrastable glasses. One such question is the role of substrate in inducing-- or suppressing-- surface freezing. This can be systematically addressed by performing rigorous free energy and/or rate calculations. 
%Also, it is interesting to study the combined effect of $\epsilon_S$ and substrate corrugation on the microstructure of the frozen monolayers. 
Finally, the ultra-thin films of long-chain alkanes can be studied to investigate the possible interplay between solid-liquid and vapor-liquid interfaces. All these topics can be the subject of further studies. 

\acknowledgements
We acknowledge the support from the National Science Foundation, NSF Grant No.~CHE-123343 (to P. G. D.) and the Princeton Center for Complex Materials, a MRSEC supported by NSF Grant No.~DMR-1420541. These calculations were performed on the Terascale Infrastructure for Groundbreaking Research in Engineering and Science (TIGRESS) at Princeton University. We gratefully acknowledge A. Panagiotopoulos, M. Ediger, D. Sussmann, M. Tylinski, Y. E. Altabet and R. Singh  for useful discussions.

\bibliographystyle{apsrev}
\bibliography{References}

\end{document}